\documentclass[final,5p,times,twocolumn,fleqn]{elsarticle}
\usepackage{amssymb,bm}
\usepackage{graphics}
\usepackage{tensor}
\usepackage{lipsum,amsmath,multicol}
\usepackage{braket,slashed}

\journal{}

\usepackage[usenames,dvipsnames]{color}
\usepackage{ulem}

\begin{document}

\begin{frontmatter}

\title{
Nondiffractive feature of $\gamma N\to \rho^\pm N$ with
$\rho$-meson electromagnetic multipole moments}

\author[KAU]{Byung-Geel Yu}
\ead{bgyu@kau.ac.kr}

\author[KAU]{Kook-Jin Kong}
\ead{kong@kau.ac.kr}

\address[KAU]{Research Institute of Basic Science, Korea Aerospace
University, Koyang, Gyeonggi 412-791, Korea}

\begin{abstract}

We investigate photoproduction of charged $\rho$ off the nucleon
using $\rho(770)+\pi(140)$ Regge pole exchanges by considering the
$\rho$-meson electromagnetic multipole moments. The significance
of the Ward identity at the $\gamma\rho\rho$ vertex is emphasized
for current conservation in the process. Given $\pi$ exchange with
the well-known coupling constants for $\gamma\pi\rho$ and $\pi
NN$, we analyze the role of the $\rho$ exchange in the $\gamma
p\to \rho^+n$ and $\gamma n\to\rho^-p$ processes without
model-dependences except for the magnetic moment
$\mu_{\rho^\pm}=\pm2.01$ and electric quadrupole moment ${\cal
Q}_{\rho^\pm}=\pm0.027$ fm$^2$ taken from theoretical estimates.
The nondiffractive feature of both cross sections is reproduced
with a rapid decrease beyond the resonance region by the dominance
of $\pi$ exchange over the $\rho$\,.
Cross sections for differential and density matrix elements are
presented to compare with existing data. The parity and photon
polarization asymmetries are predicted to demonstrate the apparent
roles of the $\rho$-meson electromagnetic multipole moments.

\end{abstract}

\end{frontmatter}

 A charged $\rho$-meson has electromagnetic (EM) multipole moments,
where the canonical values for the magnetic dipole and electric
quadrupole moments are expected to be $\mu_\rho=e_\rho/m_\rho$ and
${\cal Q}_\rho=-e_\rho/m^2_\rho$ in the limit of a point-like
particle \cite{brodsky}. Thus, it is interesting to investigate
photoproduction of charged $\rho$ \cite{clark},
\begin{eqnarray}
&&\gamma+p\to \rho^++n,\\
&&\gamma+n\to \rho^-+p,
\end{eqnarray}
because they provide a testing ground for observing any deviations in
the $\rho$-meson EM multipole moments from the canonical values
due to the substructure of the vector meson.

Unlike much studied case of the $\rho^0$ process featuring
diffractive Pomeron exchange \cite{laget2001}, little is known
about the production mechanism in the charged process.

In experimental studies, cross sections were measured for the
total, differential, and spin density matrix elements of the
$\gamma p\to \rho^+n$ process in the energy range of $E_\gamma$ =
2.8 $\sim$ 4.8 GeV by the LAMP2 group \cite{barber}, and at
$E_\gamma$ = 9.6 GeV by the Rochester-Cornell collaboration
\cite{abramson}. In the case of the $\gamma n\to \rho^-p$ process,
the total and differential cross sections in the range $E_\gamma$
= 1$\sim$5 GeV were extracted from the deuteron target by the
ABBHHM collaboration at DESY \cite{hilpert,benz}. In particular,
the data obtained from the latter process had a sharp peak in the
cross section, with $\sigma_{max}\approx$ 7 $\mu$b at around
$E_\gamma\approx$ 1.6 GeV and a steep decrease beyond the
resonance region in a similar manner to the former process. Hence,
both cross sections exhibit the typical features of a
nondiffractive process, where photoproduction of the charged
$\rho$ is expected to proceed via nonresonant meson exchanges. We
note that such a rapid decrease in the cross section over the peak
with respect to the photon energy has also been observed in
$\gamma p\to\rho^-\Delta^{++}$ \cite{barber} and $\gamma p\to
K^*\Lambda$ photoproduction measured recently by the JLab $~{}$
CLAS collaboration \cite{tang}.

In this work, we investigate charged processes $\gamma p\to
\rho^+n$ and $\gamma n\to \rho^-p$ within the Regge framework for
$\rho+\pi$ exchanges, where we utilize the Born approximation
amplitude for the gauge invariance of the $t$-channel $\rho$
exchange. Existing data regarding the cross sections, including
spin density matrix elements, are analyzed without fit parameters
because we use no cutoff function for the Regge pole exchange.
Moreover, the coupling constants related to the $t$-channel $\rho$
and $\pi$ exchanges are known from the decay width as well as
another reaction process, i.e., pion photoproduction, so the role
of $\rho$ with the EM multipole moments can be clarified without
model dependence.

However, previous studies of this issue were affected by the
theoretical limits due to the difficulty establishing gauge
invariance for the exchange of charged $\rho$ with the
$\gamma\rho\rho$ vertex coupling to charge, magnetic dipole, and
electric quadrupole moments \cite{clark,joos,reka}.
We recall that an on-shell vertex for the $\gamma\rho\rho$
coupling employed in these studies cannot satisfy the Ward identity,
which should be addressed prior to any type of prescription for
current conservation \cite{zeppen,gross}.
In addition, a simplified charge-coupling of the $\gamma\rho\rho$
vertex was considered previously in the Regge model for the
$\gamma N\to\rho^\pm N$ process \cite{laget2011}. By satisfying
gauge invariance, the model obtained agreement with existing data.
\\

\begin{figure}[t]
\centering
\includegraphics[width=4cm]{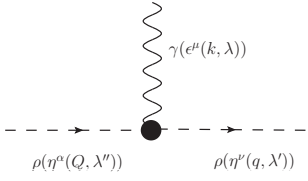}
\caption[]{$\gamma\rho\rho$ vertex for the incoming and
outgoing $\rho$ of the momenta and polarizations $(Q\,,\lambda'')$ and
$(q,\,\lambda')$. } \label{fig1}
\end{figure}

As depicted in Fig. \ref{fig1}, the divergence of the
$\gamma\rho\rho$ vertex should respect the Ward identity
\cite{zeppen},
\begin{eqnarray}\label{wardid}
k_\mu\Gamma^{\mu\nu\alpha}_{\gamma
\rho\rho}(q,Q)={(D^{-1})}^{\nu\alpha}(q)-{(D^{-1})}^{\nu\alpha}(Q)
\end{eqnarray}
with respect to the propagator chosen in the unitary gauge,
\begin{eqnarray}
D^{\nu\alpha}(q)={-g^{\nu\alpha}+q^\nu q^\alpha/ m_\rho^2\over
q^2-m^2_\rho}
\end{eqnarray}
and its inverse
\begin{eqnarray}\label{inverse}
(D^{\nu\alpha})^{-1}(q)=(q^2-m^2_\rho)\left(g^{\nu\alpha}-{q^\nu
q^\alpha\over q^2}\right)-m^2_\rho{q^\nu q^\alpha\over q^2}\,.
\end{eqnarray}

The identity in Eq. (\ref{wardid}) allows us to derive the
charge coupling term in the
$\Gamma^{\mu\nu\alpha}_{\gamma\rho\rho}$ vertex, i.e.,
\begin{eqnarray}\label{gvv}
&&e_\rho\eta_\nu^* \Gamma^{\mu\nu\alpha}_{\gamma \rho\rho}(q,Q;k)
\eta_\alpha\epsilon_\mu\nonumber\\&&
=-\eta_\nu^*(q)e_\rho\biggl\{\left[(q+Q)^\mu g^{\nu\alpha}-Q^\nu
g^{\mu\alpha}-q^\alpha g^{\mu\nu}\right]
\nonumber\\&&\hspace{0.5cm} +\kappa_\rho(k^\nu g^{\mu\alpha}-
k^\alpha g^{\mu\nu})
-{(\lambda_\rho+\kappa_\rho)\over 2m^2_\rho} \biggl[(q+Q)^\mu
k^\nu k^\alpha \nonumber\\&&\hspace{1cm}-{1\over2}(q+Q)\cdot
k(k^\nu g^{\mu\alpha}+k^\alpha g^{\mu\nu})
\biggr]\biggr\}\eta_\alpha(Q) \epsilon_\mu(k)\,,
\end{eqnarray}
which leads to the conservation of current in a rather simple
manner, which is similar to the case of pion photoproduction with
the contact term. In this case, $\epsilon^\mu(k)$ and
$\eta^\nu(q)$ are polarizations of the photon and $\rho$-meson for
momenta $k$ and $q$, respectively. The $\rho$-meson magnetic
moment with the anomalous magnetic moment $\kappa_\rho$ is taken
from Lee and Yang \cite{yang}, and the electric quadrupole moment
$\lambda_\rho$ is taken to be gauge-invariant itself according to
Gross and Riska \cite{gross}. For on-shell $\gamma\rho\rho$
coupling, i.e., $\eta^*\cdot q=0$, they are reduced to
$\mu_{\rho}=(1+\kappa_\rho){e_\rho\over 2m_\rho}$ and ${\cal
Q}_\rho=\lambda_\rho{e_\rho\over m^2_\rho}$ with $\kappa_\rho= 1$
and $\lambda_\rho =-1$ for a point-like $\rho$ meson
\cite{brodsky}, as stated earlier. In this study, we select
$\kappa_\rho=1.01$ in favor of its naturalness and
$\lambda_\rho=-0.41$ in accordance with ${\cal Q}_\rho=-0.027$
fm$^2$ \cite{maris}.

\begin{figure}[t]
\centering
\includegraphics[width=7cm]{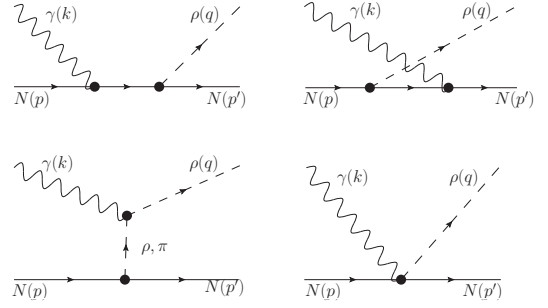}
\caption[]{Feynman diagrams for $\gamma N\to\rho^\pm N$. Nucleon
pole terms in the $s$- and $u$-channels are necessary together with the contact term
for the gauge invariance of the $t$-channel $\rho$
exchange. } \label{fig2}
\end{figure}

We now consider the production amplitude for the $\gamma
N\to\rho^\pm N$ process based on the diagrams shown in Fig.
\ref{fig2}. Considering the parity and decay channels predicted in
previous studies, the fully accounted amplitude will comprise the
scalar meson $a_0(980)$, tensor meson $a_2(1320)$, and axial
mesons $a_1(1260)$ and $b_1(1235)$, in addition to $\rho$ and
$\pi$ exchanges in the $t$-channel. For simplicity and clarity, we
consider the model of $\rho+\pi$ exchanges, although the remainder
have minor roles with contributions at an order of $10^{-2}$ at
most according to the coupling constants predicted in Refs.
\cite{bgyu,ishi,erkol}, as well as based on the analysis of the
Regge-pole fit to the data \cite{mclark}.

Then, in order to obtain a fair description of existing data up to
$E_\gamma\simeq$ 10 GeV, it is natural to employ $G$-parity
counting to determine the phases of the degenerate trajectories
$\rho$-$a_2$ and $\pi$-$b_1$ in a similar manner to pion
photoproduction \cite{glv,bgyu}.
To identify the sign of the photon-meson coupling, it
is convenient to define the $G$-parity for a photon in the case where the
isoscalar part of the photon is $G$-parity negative and the
isovector part is $G$-parity positive \cite{lipkin}.
In the diagrams of the $t$-channel exchanges shown in Fig.\ref{fig2}, the
$G$-parity negative $\pi$ and $G$-parity positive $\rho$ couple with
the isoscalar photon, 
which does not change sign in the $\gamma\pi\rho^\pm$ vertex,
whereas the $G$-parity
positive $b_1$ and $\rho$ couple to the isovector photon, 
which changes sign in the $\gamma b_1\rho^\pm$.
Therefore, the signs of the exchange-degenerate (EXD) mesons in
the production amplitudes can be written as
\begin{eqnarray}\label{regge-amp}
&&{\cal M}(\gamma N\to \rho^\pm N)\propto\left(\pm \rho+
a_2\right)+\left(\pi\pm b_1\right)\,, \nonumber\\&&\hspace{2.3cm}
\propto \left\{\begin{array}{c}\rho\, e^{-i\pi\alpha_\rho(t)}+\pi
e^{-i\pi\alpha_\pi(t)}\\-\rho(-1)+\pi(1)\end{array}\right\}+\cdots,
\hspace{0.5cm}
\end{eqnarray}
where the respective phases are determined according to the
addition of each canonical phase,
${1\over2}((-1)^J+e^{-i\pi\alpha_J(t)})$. The $\rho$ exchange
with signs denoting the $\rho$-meson charge $e_\rho$ represents
the gauge-invariant amplitude ${\cal M}_{\rho^\pm N}$, as given
in Eqs. (\ref{rho+}) and (\ref{rho-}) in the following.

In numerical calculations, either the complex or
constant phase for the $\rho$ exchange will reproduce a slow
decrease in the cross section according to the energy-dependence,
\begin{eqnarray}
\sigma\propto s^{\alpha(0)-1},
\end{eqnarray}
which contradicts the rapid slope in the decrease observed in
both cross sections. Furthermore, the $\rho+\pi$ exchanges with
the EXD phases in Eq. (\ref{regge-amp}) are largely overestimated
for both cross sections, unless the intercept of the $\rho$ trajectory is
smaller than usual, e.g., $\alpha_\rho(t)=0.8\,t+0.35$. Therefore, to obtain
agreement with the experiments, we must use the cutoff
function with the trajectory $\alpha_\rho(t)=0.8\,t+0.55$, as
conventionally adopted by Laget \cite{laget2011}, or we assume a
violation of the EXD in one of these $\rho$-$a_2$ and $\pi$-$b_1$
pairs, as employed by Clark and Donnachie \cite{mclark}.
To favor the dominance of $\pi$
exchange with the EXD phase over $\rho$, we
break the EXD of the $\rho$-$a_2$ pair to assign the canonical
phase, ${1\over2}(-1+e^{-i\pi\alpha_\rho})$ to $\rho$ with the
trajectories for $\rho$ and $\pi$ employed in previous studies
\cite{mclark,clark3},
\begin{eqnarray}
&&\alpha_\rho(t)=0.9\,(t-m_\rho^2)+1,\\
&&\alpha_\pi(t)=0.7\,(t-m_\pi^2),
\end{eqnarray}
which are in better agreement with the total cross section data.

Given the Regge pole,
\begin{eqnarray}
{\cal
R}^{\varphi}(s,t)={\pi\alpha'_J\over\Gamma[\alpha_J(t)+1-J]\sin[\pi\alpha_J(t)]}
\left({s\over s_0}\right)^{\alpha_J(t)-J}\,,\ \ \ \ \ \
\end{eqnarray}
written collectively for a meson $\varphi$ of spin-$J$, the
conserved $\rho$-exchange for $\gamma p\to\rho^+n$ is
given by
\begin{eqnarray}\label{rho+}
&&{\cal M}_{\rho^+n}=\sqrt{2}\,\bar{u}(p')\eta^*_{\nu}(q)\biggl(
\Gamma_{\rho pn}^\nu(q){\rlap{\,/}p+\rlap{/}k+M_p\over
s-M_p^2}\Gamma^\mu_{\gamma
pp}(k)\nonumber\\&&+e\Gamma_{\gamma\rho\rho}^{\mu\nu\alpha}(q,Q)D_{\alpha\beta}(Q)
\Gamma^\beta_{\rho pn}(Q)
-e{g_\rho^t\over4M_p}[\gamma^\nu,\gamma^\mu]
\biggr)\epsilon_{\mu}(k)u(p)\nonumber\\&&\times\,(t-m_\rho^2)\times{\cal
R}^\rho(s,t)\times
{1\over2}\left(-1+e^{-i\pi\alpha_\rho(t)}\right),
\end{eqnarray}
and for $\gamma n\to\rho^- p$,
\begin{eqnarray}\label{rho-}
&&{\cal
M}_{\rho^-p}=\sqrt{2}\,\bar{u}(p')\eta^*_{\nu}(q)\biggl(\Gamma^\mu_{\gamma
pp}(k){\rlap{\,/}p'-\rlap{/}k+M_p\over u-M_p^2}\Gamma_{\rho
pn}^\nu(q) \nonumber\\
&&-e\Gamma_{\gamma\rho\rho}^{\mu\nu\alpha}(q,Q)D_{\alpha\beta}(Q)
\Gamma^\beta_{\rho np}(Q)
+e{g_\rho^t\over 4M}[\gamma^\nu,\gamma^\mu]
\biggr)\epsilon_{\mu}(k)u(p)\nonumber\\&&\times\,(t-m_\rho^2)\times{\cal
R}^\rho(s,t)\times
{1\over2}\left(-1+e^{-i\pi\alpha_\rho(t)}\right),
\end{eqnarray}
respectively. The coupling vertices $\gamma NN$ and $\rho NN$ are
given by
\begin{eqnarray}
&&\Gamma^\nu_{\rho NN}(q)=g^v_{\rho}\gamma^\nu+{g^t_{\rho}\over
4M}[\gamma^\nu,\rlap{\,/}q],\\
&&\Gamma^\mu_{\gamma pp}(k)=e\gamma^\mu-{e\kappa_p\over
4M}[\gamma^\mu,\rlap{\,/}k],
\end{eqnarray}
where $g^v_\rho$=2.6 and $g_\rho^t$=9.62 are taken from the vector
dominance with the ratio of $g_\rho^t/g_\rho^v=3.7$, which is
consistent with the nucleon's anomalous magnetic moment. For
Reggeization, we introduce the nucleon pole term for the gauge
invariance of the $t$-channel vector meson exchange. Given the
significance of the magnetic interaction of the spin-1 particle,
we preserve the nucleon magnetic moment $\kappa_p=1.79$ in the
$s$-channel for the $\gamma p\to \rho^+n$ process, and in the
$u$-channel proton pole term for $\gamma n\to \rho^-p$.

\begin{figure}[t]
\includegraphics[width=8cm]{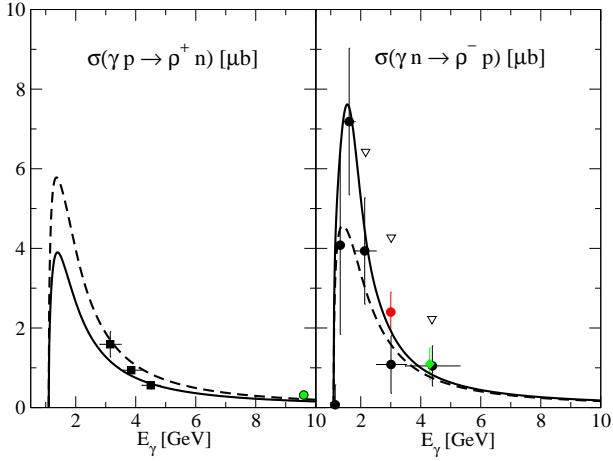}
\caption[]{ Dependence of the total cross section on
the sign of the $\gamma\pi\rho$ coupling for $\rho^+$ (left) and
$\rho^-$ (right) processes. Solid lines represent results based on
$g_{\gamma\pi\rho}$ = $-0.224$ for $\pi$ exchange plus the
$\rho$ exchange with $\kappa_{\rho}$ = 1.01 and $\lambda_\rho$ =
$-0.41$. Dashed lines denote the case where $g_{\gamma\pi\rho}$ =
$+0.224$. The data are from previous studies
\cite{barber,abramson,hilpert,benz}} \label{fig3}
\end{figure}

For the $\pi$ exchange, we write the Regge-pole amplitude as
\begin{eqnarray}\label{regge02}
&&\mathcal{M}_{\pi^\pm}=i\sqrt{2}\,{g_{\gamma\pi\rho}\over
m_0}g_{\pi NN}
\varepsilon^{\mu\nu\alpha\beta}\epsilon_\mu\eta_\nu^* k_\alpha
q_\beta\,\bar{u}(p')\gamma_5\,u(p)\nonumber\\&&\hspace{1cm}\times\,{\cal
R}^\pi(s,t) \times
\left\{\begin{array}{c}e^{-i\pi\alpha_\pi(t)}\\1\end{array}\right\},
\end{eqnarray}
for $\gamma p\to \rho^+n$ (upper), and $\gamma n\to \rho^-p$
(lower) with $g_{\pi NN}=13.4$ and the mass parameter $m_0=1$ GeV.
Then, the coupling constant $|g_{\gamma\pi\rho}|=0.224$ is
estimated from the width $\Gamma_{\rho^\pm\to\pi\gamma}=68$ keV,
and we take the sign of the $\pi$ contribution with
${\gamma\pi\rho}$ coupling relative to $\rho$ in order to achieve
more consistency with existing data related to both processes.

Figure \ref{fig3} shows the dependence of the total cross sections on
the sign of the $g_{\gamma\pi\rho}$ coupling constant for the $\rho^+$ and
for $\rho^-$ processes. We recall that the charge
asymmetry of the charged $\rho$ photoproduction off the deuteron
target was measured by Abramson et al. \cite{abramson} as
\begin{eqnarray}
{\sigma_{\gamma d\to \rho^+m}-\sigma_{\gamma d\to
\rho^-m}\over\sigma_{\gamma d\to \rho^+m}+\sigma_{\gamma d\to
\rho^-m}}\approx-0.11\pm0.03,
\end{eqnarray}
based on the average of the invariant mass $M_{\pi^+\pi^-}$ interval.
Thus, we can determine the ratio $\sigma_{\gamma d\to
\rho^-m}/\sigma_{\gamma d\to \rho^+m}\approx 1.25$ to predict that
$\sigma_{\rho^-}$ is larger than $\sigma_{\rho^+}$ in Fig.
{\ref{fig3}}. This is true for the ratio of the total cross
sections between $\pi^+$ and $\pi^-$ processes, which have the same
isospin structure as $\rho^+$ and $\rho^-$ processes. Thus, the
$g_{\gamma\pi\rho}$ positive case is discarded.

\begin{figure}[]
\includegraphics[width=8cm]{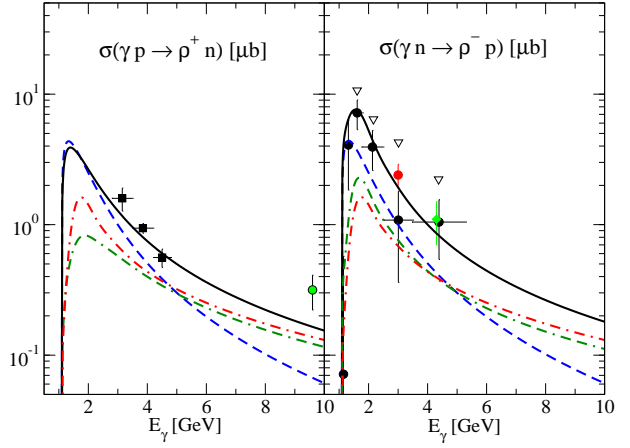}
\caption[]{Contributions of $\rho$ and $\pi$ exchanges to the total
cross section for $\rho^+$ and for $\rho^- $ processes. The
dashed line (blue) is due to $\pi$ exchange and the
dash-dotted line (red) is based on the single $\rho$ exchange. The
dash-dash-dotted line (green) is obtained from the gauge-invariant $\rho$
exchange, ${\cal M}_{\rho^\pm N}$. } \label{fig4}
\end{figure}

\begin{figure}[t]
\includegraphics[width=8cm]{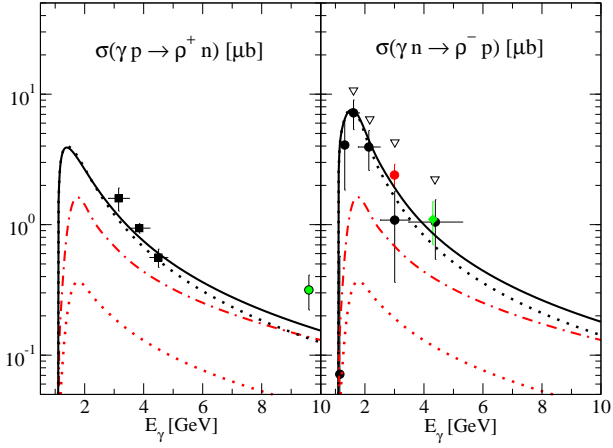}
\caption{Dependence of the total cross section on the EM multipole
moments of the $\rho$-meson. The red dotted line is based on the
single $\rho$ exchange with $\kappa_\rho=0$ and $\lambda_\rho=0$.
The black dotted line is the resulting $\sigma$ obtained for the $\rho^+$
process (left), and similarly for the $\rho^-$ process (right).}
\bigskip
\bigskip
\label{fig5}
\end{figure}

\begin{figure}[t]{}
\includegraphics[width=8cm]{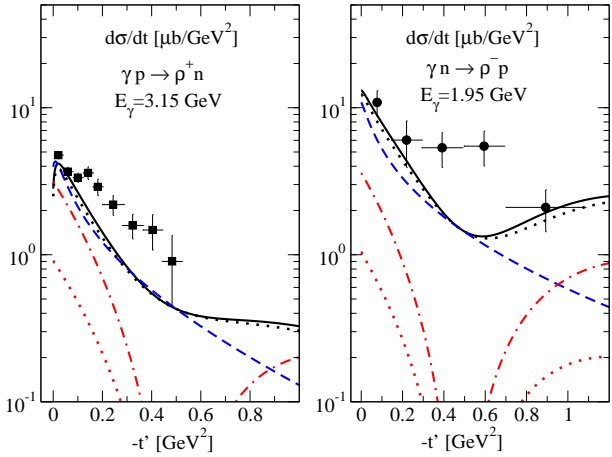}
\caption[]{Contributions of $\pi$ and $\rho$ exchanges to the
differential cross section for $\rho^+$ at $E_\gamma=3.15$ GeV and
for $\rho^- $ at $E_\gamma=1.95$ GeV. The notations are the same as those used in
Figs. \ref{fig4} and \ref{fig5}. The data are from previous studies
\cite{barber,benz}.} \label{fig6}
\end{figure}
\begin{figure}[]
\centering
\includegraphics[width=6.2cm]{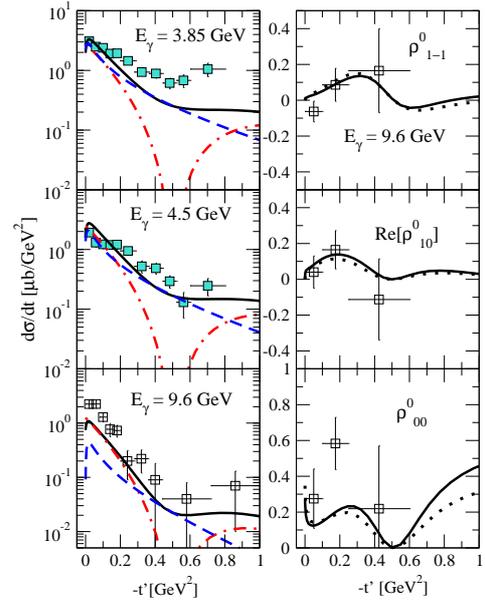}
\caption{Differential cross sections in three energy bins and
density matrix elements $\rho^0_{\lambda\lambda'}$ at $E_\gamma$ =
9.6 GeV for $\gamma p\to\rho^+n$. The notations are the same as those used in Fig.
\ref{fig6}. Data for the full-square are from Barber et al. \cite{barber} and those for the
empty-square are from Abramson et al. \cite{abramson}.} \label{fig7}
\end{figure}

Figure \ref{fig4} shows the contributions of $\rho$ and $\pi$ exchanges.
The dominance of the latter exchange before
$E_\gamma\simeq$ 5 GeV in both processes is responsible for the
steep decrease in both cross sections. As shown in the figure, the different
contributions of $|\rho-\pi|^2$ ($\rho$ denotes ${\cal
M}_{\rho^\pm N}$ on the green line) for $\rho^+$ and $|-\rho-\pi|^2$
in $\rho^-$ production via the negative coupling $\gamma\pi\rho$
can help us to understand the relative size of the cross
sections in this region. Above $E_\gamma\simeq 5$ GeV, the
contribution of $\rho$ exchange dominates that of
$\pi$, as shown by the experimental data obtained at $E_\gamma=9.6$ GeV
\cite{abramson}.

The roles of the EM multipole moments of $\rho$ exchange are
illustrated in Fig. \ref{fig5} in order to show the contributions of
$\kappa_\rho$ and $\lambda_\rho$ to the cross section. The single
$\rho$ exchange (the $t$-channel exchange in Fig.
\ref{fig2}) with and without these contributions differs by an
order of magnitude, as shown by the dash-dotted and dotted lines.
The role of $\kappa_\rho$ is more significant than that of
$\lambda_\rho$.

The differential cross sections for $\rho^+$ and $\rho^-$
processes are presented in Fig. \ref{fig6}, which exhibit a dipped
structure for $\rho$ exchange at $-t\approx 0.5$ GeV$^2/c^2$ due
to the nonsense-wrong-signature-zero of the $\rho$ trajectory,
i.e., $\alpha_\rho(t)=0$, from the canonical phase. This feature
makes the angular distribution increasing over $-t\approx 0.5$
GeV$^2/c^2$, which is apparent in the differential cross sections
of $\rho^+$ in other energy ranges, as shown in Fig. \ref{fig7}.
However, we need more data to confirm such a structure for the
$\rho^-$ process. We note that our estimate of the contribution of
the $\pi$ exchange to $d\sigma/dt$ at $E_\gamma=9.6$ GeV in the
$\rho^+$ process is similar to that given by Abramson et al.
\cite{abramson}, but different from that provided by Benz et al.
\cite{benz} at $E_\gamma=1.95$ GeV in the $\rho^-$ process.

Figure \ref{fig7} shows the differential cross sections together
with the density matrix elements $\rho^0_{\lambda\lambda'}$
estimated in the Gottfried-Jackson frame \cite{schilling,zhao} for
the unpolarized $\rho^+$ process. Since the latter observables
relate the spin polarization of the final vector meson to that of
the initial photon in the helicity amplitude, the spin
correlations involved in the production mechanism provide a
further test of the validity of the model predictions. In our
proposed framework, the density matrix elements are reproduced by
the oscillatory behavior of the $\rho$ exchange due to the
canonical phase and thus they agree with the data.

\begin{figure}[]
\centering
\includegraphics[width=8cm]{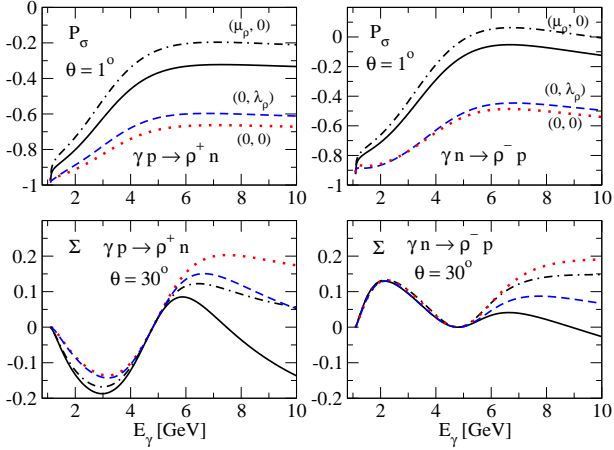}
\caption{Energy dependence of the parity asymmetry $P_\sigma$ and
photon polarization asymmetry $\Sigma$. The labels on the curves
denote whether the EM multipole moments are turned on and off. }
\label{fig8}
\end{figure}

Finally, we present model predictions for the parity asymmetry
\cite{schilling} $P_\sigma=2\rho^1_{1-1}-\rho^1_{00}$ at
$\theta=1^\circ$ in the c.m. frame and the photon polarization
asymmetry \cite{zhao} $\Sigma={2\rho^1_{11}}+\rho^1_{00}$ at
$\theta=30^\circ$ in Fig. \ref{fig8}. The role of the $\rho$-meson
EM multipole moments is apparent in these observables, so the
measurements of $P_\sigma$ and $\Sigma$ could be useful for
guiding the determination of $\mu_\rho$ and $\lambda_\rho$ in
experiments.

In summary, we investigated photoproduction of charged $\rho$ for
positive and negative cases based on the Regge model, where the
$\gamma\rho\rho$ vertex fully accounted for the magnetic dipole
and electric quadrupole moments. We stressed the validity of the
Ward identity of the $\gamma\rho\rho$ vertex for the gauge
invariance of the process. Using the non-degenerate phase for the
$\rho$ exchange based on empirical evidence for a steep decrease
in the cross sections, we showed that these processes are
dominated by $\pi$ exchange over $\rho$, thereby demonstrating the
nondiffractive feature of the processes, as well as the dipped
structure in the differential cross section and the oscillatory
behavior of the density matrix. We investigated the role of the
$\rho$-meson EM multipole moments $\mu_{\rho^\pm}=\pm2.01$ and
${\cal Q}_{\rho^\pm}=\pm0.027$ fm$^2$ to analyze the production
mechanism based on comparisons with existing data. We predicted
the parity and photon polarization asymmetries within the present
framework to facilitate future experimental observations. In
particular, measurements of cross sections for the $\gamma
p\to\rho^+n$ and $\gamma n\to\rho^-p$ processes in the resonance
region are desirable to allow further development of the theory of
photoproduction of charged vector meson.

        \section*{Acknowledgments}
We are grateful to Ho-Meoyng Choi for fruitful discussions. This
work was supported by the grant NRF-2013R1A1A2010504 from the
National Research Foundation (NRF) of Korea.

\end{document}